\documentclass[runningheads]{llncs}
\usepackage[T1]{fontenc}
\usepackage{graphicx}
\usepackage{xspace}
\usepackage{url}
\usepackage{hyperref}
\usepackage{xcolor}
\usepackage{tikz}
\usepackage{paralist}
\usepackage{listings}
\usetikzlibrary{shapes.geometric,backgrounds,calc,positioning,fit,shapes,arrows,arrows.meta}
\usepackage{fancybox}
\usepackage{calc}
\usepackage[english]{babel}
\newcommand{\asmeta}{\texttt{Asmeta}\xspace}%
\newcommand{\asmetaL}{{\tt AsmetaL}\xspace}
\newcommand{\avalla}{{\tt Avalla}\xspace}
\newcommand{\asmetaSMV}{{\tt AsmetaSMV}\xspace}

\lstdefinelanguage{AsmetaL} 
{morekeywords={module, par, endpar, if, endif, then, else, seq, endseq, 
signature, definitions, asm, import, function, domain, main, rule, macro, 
invariant, over, choose, let, endlet, with, ifnone, forall, abstract, default, 
init, do, agent, dynamic, controlled, monitored, in, out, static, derived, 
subsetof, switch, case, endswitch, enum, CTLSPEC, LTLSPEC, JUSTICE, INVARSPEC, NAME, next, AG, AX, implies},
	sensitive=true, morecomment=[l]{//}, morecomment=[s]{/*}{*/},
	morecomment=[l][\color{white}\tiny]{'},
	morestring=[b]",tabsize=1, columns=fullflexible, basicstyle=\scriptsize\sffamily, captionpos=b, frame=lines}
\lstdefinelanguage{Avalla} 
{morekeywords={load,step, set, check,scenario, pick, in, or},
	sensitive=true, morecomment=[l]{//}, morecomment=[s]{/*}{*/},
	morecomment=[l][\color{white}\tiny]{'},
	morestring=[b]",tabsize=3, columns=fullflexible, basicstyle=\scriptsize\sffamily, 
	captionpos=b,escapechar=|,frame=lines}

\begin{document}
\title{Writing properties in \asmeta with AI} 
\title{Using AI to formalize and validate properties in \asmeta}
\title{Formalizing and validating properties in \asmeta with Large Language Models (Extended Abstract)}
%
%
\author{Andrea Bombarda\inst{1}\orcidID{0000-0003-4244-9319} \and
Silvia Bonfanti\inst{1}\orcidID{0000-0001-9679-4551}\and
Angelo Gargantini\inst{1}\orcidID{0000-0002-4035-0131} \and
Nico Pellegrinelli\inst{1}\orcidID{0009-0000-4944-6845}}

\authorrunning{A. Bombarda et al.}

\institute{University of Bergamo, Bergamo, Italy
\email{\{andrea.bombarda, silvia.bonfanti, angelo.gargantini, nico.pellegrinelli\}@unibg.it}}
\maketitle              
\begin{abstract}
Writing temporal logic properties is often a challenging task for users of model-based development frameworks, particularly when translating informal requirements into formal specifications. In this paper, we explore the idea of integrating Large Language Models (LLMs) into the \asmeta framework to assist users during the definition, formalization, explanation, and validation of temporal properties. We present a workflow in which an LLM-based agent supports these activities by leveraging the \asmeta specification and the feedback produced by the model checker. This work serves as a proof of concept that illustrates the feasibility and potential benefits of such an integration through representative examples.

\end{abstract}
%
%
%

\section{Introduction}
Modern software and system engineering projects are increasingly complex, often involving heterogeneous components, tight safety constraints, and short development cycles. Traditional development approaches, which rely heavily on manual coding and late-stage testing, struggle to ensure early validation, traceability, and consistency between design and implementation. This leads to higher costs and risks, especially in safety-critical domains such as automotive, aerospace, and healthcare. 
Model-Based Development (MBD) offers an effective alternative. By relying on executable models, it enables early simulation and verification of system behavior, reduces ambiguity in requirements through formal or semi-formal representations, supports automatic code generation to ensure alignment between design and implementation, and enhances traceability and compliance with industry standards.
A variety of tools support MBD, and, in this work, we focus on the \asmeta framework~\cite{Arcaini2021a,Bombarda2024}.
The framework allows developers to work with Abstract State Machines (ASMs)~\cite{Brger2003}, which extend Finite State Machines (FSMs) by replacing unstructured control states with states that can contain arbitrarily complex data.

\section{Objectives} \label{sec:objective}
The \asmeta framework provides a suite of tools that support developers throughout the software lifecycle, including design, development, and operation. In this work, we concentrate on the design phase, and in particular on verification activities, which are carried out using the model checker integrated into \asmetaSMV~\cite{Arcaini2010}.

Starting from the functional requirements, \asmeta enables users to model a system using the \asmetaL language. Users can embed Computation Tree Logic (CTL) and Linear Temporal Logic (LTL) properties directly within the ASM model, expressing them over the \asmeta signature.
The \asmetaSMV tool, then, automatically translates the ASM model (including both the system model and the LTL/CTL properties) into a model of the symbolic model checker NuSMV~\cite{Cimatti2002}, which is used for verification.

Based on feedback collected from \asmeta users, particularly students enrolled in courses where \asmeta is taught, we observed that many of them encounter difficulties when writing LTL and CTL properties. While formulating properties in natural language is generally straightforward, challenges arise when translating these descriptions into the corresponding \asmetaL expressions.

To address this issue, we propose integrating Large Language Models (LLMs) into the workflow.
We have identified, after analyzing existing works, at least four ways in which LLMs can assist the designer in the verification of TL properties.

\paragraph{\textbf{O1}. Assisting during the definition of the property in natural language:}

Given an \asmeta specification, LLMs can be leveraged to interpret the domain context and automatically derive properties, expressed in natural language, that are valid within that context. 
These properties are neither intended to be exhaustive nor universally guaranteed; rather, they serve to improve the user’s understanding of the domain and to potentially reveal constraints or assumptions that were implicit in the original requirements. There are already some experiments proposing Retrieval-Augmented Property Generation in other contexts, like PropertyGPT~\cite{Liu_2025}.

\paragraph{\textbf{O2}. Helping the designer to write the corresponding LTL/CTL properties:}
Given an \asmeta specification that includes the system behavior and properties expressed in natural language, the LLM automatically translates these properties into their corresponding LTL or CTL formulations within the \asmeta specification. By doing so, we aim to reduce the gap between informal requirement descriptions and formal verification artifacts, ultimately improving usability and lowering the entry barrier for new users.

Furthermore, the generated properties must be formulated with awareness of, and consistency with, the functions defined in the \asmeta specification, thereby ensuring that the resulting LTL/CTL formulas are consistent with the operational semantics of the model. This implies that the LLM must be context-aware: it must recognize the roles and types of functions, and incorporate these elements appropriately into temporal operators.

\paragraph{\textbf{O3}. Explanation of the TL properties using natural language:}

Formalizing requirements enables systematic reasoning about inconsistencies, the detection of ambiguities, and the identification of critical issues in system models. Temporal logic formulae are a natural choice for specifying requirements related to desired system behaviors. However, understanding and effectively using temporal logic demands a strong formal background. Therefore, there is a need for approaches that make temporal logic formulae more interpretable for engineers, domain experts, and other stakeholders involved in the development process. For this goal, we assume that LLMs can be used to translate TL properties back to natural language. For this scope, we plan to adopt (or adapt, if necessary) the technique presented in \cite{Cherukuri22}.

\paragraph{\textbf{O4}. Validation of the TL properties using witnesses and counterexamples:}

When a property fails verification, \asmetaSMV produces a counterexample that can be automatically exported as an \avalla scenario.
Likewise, successful executions could be exported as scenarios that highlight witnesses.
Large Language Models (LLMs) can support developers’ understanding by translating \avalla scenarios into natural language, thereby clarifying their meaning and describing the behavior they implement.


\medskip

We report on preliminary investigations focused on outlining the process to be adopted for integrating an LLM-based assistant into the \asmeta framework. The goal at this stage is to define the process required to incorporate an LLM-based assistant into the \asmeta framework.

\section{Methodology and Walkthrough Demonstration}

This work follows a multi-step methodology aimed at designing, integrating, and empirically evaluating an LLM-based assistant for the automatic generation of temporal properties in \asmeta specifications. Since this is a preliminary study, our focus is on defining the overall process, identifying key challenges, and conducting initial feasibility analyses rather than delivering a fully operational framework.

\begin{figure}[tb]
    \centering
    \begin{tikzpicture}[
    node distance=.7cm and 1cm,
    box/.style={rectangle, text=black, rounded corners, minimum width=2.5cm, minimum height=.7cm, align=center,draw=black},
    imagebox/.style={rectangle, text=black, rounded corners, minimum width=1.5cm, minimum height=1cm, align=right},
    line/.style={-{Latex[length=3mm]}, thick, black}
]

\node[box] (planning) {LLM-Based Agent};
\node[box, right=of planning, xshift=2cm] (debug) {\asmetaSMV};
\node[above left=7mm and -15mm of planning] (input-text) {\asmeta specification};
\node[above right=7mm and -5mm of planning] (prompt-text) {Prompt};
\node[below=of planning] (output) {Enriched \asmeta specification};

\draw[line] (input-text) -- (planning);
\draw[line] (prompt-text) -- (planning);
\draw[line] (planning) -- (output);
\draw[line,dashed] ([yshift=-2mm]planning.east) -- ([yshift=-2mm]debug.west);
\draw[line] ([yshift=2mm]debug.west) -- ([yshift=2mm]planning.east);

\end{tikzpicture}
    \caption{Our solution workflow.}
    \label{fig:workflow}
\end{figure}
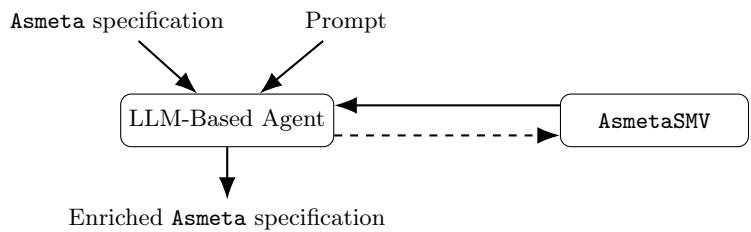

The LLM-based assistant, whose workflow is reported in Figure~\ref{fig:workflow}, should be able to generate and explain LTL and CTL properties that are not only syntactically correct, but also meaningful with respect to the given \asmeta model. This implies that the assistant must interpret the signature of the specification, understand the types of functions, and generate formulas or explanations that consistently refer to elements defined by the model. 

The user begins by providing both an \asmeta specification and the informal requirements of properties to be formalized, within a specifically designed prompt. The assistant, then, extracts relevant signatures from the \asmeta specification and translates the requirements into an LTL or CTL formula embedded into the \asmeta specification, or provides explanations complying with the requested tasks and using the same functions as in the \asmeta specification.

We expect users to write requirements by following the EARS (Easy Approach to Requirements Syntax)~\cite{5328509}, a structured way to write requirements so they are clear, consistent, and less ambiguous.

In the following, we describe how we envision that our approach would help in each of the four objectives described in Section~\ref{sec:objective}. Our demonstration is based on direct interactions with the ChatGPT 5.2 Instant model and aims at checking the feasibility of our approach.

Let us consider the \asmeta specification of a simple clock as in Listing~\ref{lst:asmetaspec}.
For the objective \textbf{O1}, users may ask the LLM-Based Agent to list the \emph{three} most important properties guaranteed by the \asmeta specification.
In this case, the LLM replies with the three following properties (each with its own description):

\smallskip
\noindent\fbox{\begin{minipage}[t]{1\columnwidth - 2\fboxsep - 2\fboxrule}%
\ttfamily
\begin{compactitem}
    \item Time variables always stay within valid clock ranges;
    \item Correct carry-over from seconds to minutes and hours;
    \item Time advances only when the signal is true.
\end{compactitem}
\end{minipage}}
\smallskip

Although some of these properties may be straightforward, providing a natural language explanation can help lower the barrier to understanding \asmeta specifications.

\begin{figure}[tb]
\begin{minipage}{.48\textwidth}
\begin{lstlisting}[language=AsmetaL]
asm Clock
import StandardLibrary
signature:
    domain Second subsetof Integer
    domain Minute subsetof Integer
    domain Hour subsetof Integer      
    monitored signal: Boolean  
    controlled sec: Second    
    controlled min: Minute    
    controlled h: Hour
definitions:    
    domain Second = {0 : 59}
\end{lstlisting}
\end{minipage} \hfill
\begin{minipage}{.49\textwidth}
\begin{lstlisting}[language=AsmetaL]
    domain Minute = {0 : 59}
    domain Hour = {0 : 23}
    macro rule r_IncMinHours = par
        if min = 59 then h := (h + 1) mod 24 endif
        min := (min + 1) mod 60 endpar
    main rule r_Main = if signal then par
        if sec = 59 then r_IncMinHours[] endif
        sec := (sec + 1) mod 60 endpar endif
default init s0:    
    function sec = 0
    function min = 0
    function h = 0
\end{lstlisting}
\end{minipage}
\vspace{-1.2em}
\begin{lstlisting}[caption={\asmeta specification of a simple Clock},label={lst:asmetaspec}]
\end{lstlisting}
\end{figure}

For objective \textbf{O2}, users may ask the LLM-Based Agent to encode in CTL a property verifying that when the \textsf{min} function reaches the value $59$, it is set to $0$ in the next state.
With this prompt, the LLM-Based Agent replies with the following CTL property: 

\smallskip
\noindent\fbox{\begin{minipage}[t]{1\columnwidth - 2\fboxsep - 2\fboxrule}%
\ttfamily
AG (min = 59 implies AX (min = 0))
\end{minipage}}
\smallskip

\begin{figure}[tb]
    \begin{minipage}{.49\columnwidth}
        \begin{lstlisting}[language=Avalla]
scenario ClockScenario
load Clock.asm
set signal := true;
step;
\end{lstlisting}
    \end{minipage} 
    \hfill
    \begin{minipage}{.49\columnwidth}
      \begin{lstlisting}[language=Avalla]
check sec = 1 and min = 0 and h = 0;
set signal := false;
step;
check sec = 1 and min = 0 and h = 0;
\end{lstlisting}  
    \end{minipage}
    \vspace{-1.2em}
    \begin{lstlisting}[caption={\avalla scenario for the Clock specification},label={lst:avallascenario}]
\end{lstlisting}
\end{figure}
A property such as the one generated by our LLM-Based Agent must contain only valid operators and functions. 
While LLMs can be fed with an \asmeta specification, it may happen that the LLM generates a syntactically wrong property. For this reason, as shown in Figure~\ref{fig:workflow}, we integrate our solution with \asmetaSMV. 
Once the property is generated, the model checker’s feedback is leveraged. If errors are detected, the model checker’s output is provided together with the prompt to correct the property formulation.

For objective \textbf{O3}, suppose that the considered \asmeta specification contains the property given before. 
Users may leverage our LLM-Based Agent to explain in natural language what the property means.
In this example, the output of our Agent is the following: 

\smallskip
\noindent\fbox{\begin{minipage}[t]{1\columnwidth - 2\fboxsep - 2\fboxrule}%
\ttfamily
In every reachable state of the system, if the value of min is 59, then in all possible next states the value of min will be 0.
\end{minipage}}
\smallskip

Finally, for the objective \textbf{O4}, if the \avalla scenario in Listing~\ref{lst:avallascenario} is available (e.g., as the output of the counterexample generation), users may be interested in having its high-level to better understand the reasons behind a possible failure.
In this case, the LLM-Based Agent provides the following explanation: 

\smallskip
\noindent\fbox{\begin{minipage}[t]{1\columnwidth - 2\fboxsep - 2\fboxrule}%
\ttfamily
The scenario verifies that the clock advances by one second when the signal is true, and stops advancing when the signal is false, preserving the current time.
\end{minipage}}



\section{Conclusion}

In this paper, we explored the idea of integrating Large Language Models into the \asmeta framework to support users in writing, understanding, and validating temporal logic properties. 
Motivated by the difficulties encountered by practitioners when translating informal requirements into LTL and CTL specifications, we outlined a conceptual workflow in which an LLM-based agent assists the designer across multiple verification-related tasks, ranging from property elicitation in natural language to the interpretation of witnesses and counterexamples.

Through simple examples, we showed that LLMs can plausibly bridge the gap between informal requirements and formal verification artifacts, while remaining consistent with the semantics of \asmeta specifications and the feedback provided by the model checker.
Future work will focus on turning this conceptual workflow into a concrete implementation within \asmeta.

%
%
%
 \bibliographystyle{splncs04}
 \bibliography{TLPropBib}
\end{document}